\documentclass[lettersize,journal]{IEEEtran}
\usepackage{amsmath,amsfonts}
\usepackage{algorithm}
\usepackage{array}
\usepackage[caption=false,font=normalsize,labelfont=sf,textfont=sf]{subfig}
\usepackage{textcomp}
\usepackage{stfloats}
\usepackage{url}
\usepackage{verbatim}
\usepackage{graphicx}
\usepackage{cite}
\usepackage{acronym}
\usepackage{glossaries}
\usepackage[noend]{algpseudocode}
\usepackage{amsmath}
\newcommand{\twopartdef}[4]
{
	\left\{
	\begin{array}{ll}
		#1 & \mbox{if } #2 \\
		#3 & \mbox{if } #4
	\end{array}
	\right.
}

\newcommand\scalemath[2]{\scalebox{#1}{\mbox{\ensuremath{\displaystyle #2}}}}

\begin{document}
	
\newacronym{ldpc}{LDPC}{Low-Density Parity-Check}
\newacronym{qc-ldpc}{QC-LDPC}{Quasy-Cyclic Low-Density Parity-check codes}
\newacronym{peg}{PEG}{Progressive Edge Growth}
\newacronym{al-peg}{AL-PEG}{Arhitecture-aware Layered Progressive Edge Growth}
\newacronym{peg-qc}{PEG-QC}{Quasy-Cyclic Progressive Edge Growth}
\newacronym{raw}{RAW}{Read After Write}
\newacronym{hue}{HUE}{Hardware Usage Efficency}
\newacronym{tsp}{TSP}{Travelling Salesman Problem}
\newacronym{mg}{MG}{Misra-Griess}
\newacronym{mmg}{MMG}{Modified Misra-Griess}
\newacronym{fec}{FEC}{Forward Error Correction}
\newacronym{bp}{BP}{Belief Propagation}
\newacronym{sp}{SP}{Sum-Product}
\newacronym{ms}{MS}{Min-Sum}
\newacronym{oms}{OMS}{Offset Min-Sum}
\newacronym{nms}{NMS}{Normalized Min-Sum}
\newacronym{map}{MAP}{Maximum A Posteriori}
\newacronym {snr}{SNR}{Signal-to-Noise Ratio}
\newacronym{scms}{SCMS}{Self Correcting Min-Sum}
\newacronym{vn}{VN}{Variable-Node}
\newacronym{cn}{CN}{Check-Node}
\newacronym{vnu}{VNU}{Variable-Node Unit}
\newacronym{cnu}{CNU}{Check-Node Unit}
\newacronym{vcn}{VCN}{Variable-Check Node Unit}
\newacronym{ap-llr}{AP-LLR}{A-Posteriori Log-Likelihood Ratios}
\newacronym{bram}{BRAM}{Block RAM}
\newacronym{fpga}{FPGA}{Field Programmable Gate Array}
\newacronym{srams}{SRAMs}{Static Random Access Memories}
\newacronym{sram}{SRAM}{Static Random Access Memory}
\newacronym{fifo}{FIFO}{First-In First-Out}
\newacronym{vns}{VNs}{Variable-Nodes}
\newacronym{cns}{CNs}{Check-Nodes}
\newacronym{vnus}{VNUs}{Variable Node Units}
\newacronym{cnus}{CNUs}{Check Node Units}
\newacronym{vcns}{VCNs}{Variable Check Node Units}
\newacronym{bs}{BS}{Barrel Shifter}
\newacronym{bss}{BS}{Barrel Shifters}
\newacronym{rom}{ROM}{Read Only Memory}
\newacronym{roms}{ROM}{Read Only Memories}
\newacronym{fer}{FER}{Frame Error Rate}
\newacronym{tedi}{TeDi}{Tool and Environment for HDL Template Design}
\newacronym{faid}{FAID}{Non-Surjective Finite Alphabet Iterative Decoder}
\newacronym{faids}{FAIDs}{Finite Alphabet Iterative Decoders}
\newacronym{ns-faids}{NS-FAIDs}{Non-surjective Finite Alphabet Iterative Decoders}
\newacronym{ns-faid}{NS-FAID}{Non-surjective Finite Alphabet Iterative Decoder}
\newacronym{bsc}{BSC}{Binary Symetric Channel}
\newacronym{awgn}{AWGN}{Additive White Gaussian Noise}

\title{Iterative Gradient Descent Decoding for Real Number LDPC Codes}
\author{\IEEEauthorblockN{Oana Boncalo, Alexandru Amaricai \\}
	\IEEEauthorblockA{
		Universitatea Politehnica Timisoara\\
		Timisoara, Romania\\
		Email: oana.boncalo@cs.upt.ro}}



\maketitle

\begin{abstract}
This paper proposes a new iterative gradient descent decoding method for real number parity codes. The proposed decoder, named Gradient Descent Symbol Update (GDSU), is used for a class of low-density parity-check (LDPC) real-number codes that can be defined with parity check matrices which are similar to those of the binary LDPC from communication standards such as WiFi (IEEE 802.11), 5G. These codes have a simple and efficient two stage encoding that is especially appealing for the real number field. The Gradient optimization based decoding has been a relatively simple and fast decoding technique for codes over finite fields. We show that the GDSU decoder outperforms the gradient descent bit-flipping (GDBF) decoder for rates $1/2$, $2/3$, and has similar decoding performance for the $3/4$ rate of the IEEE 802.11 codes standard.

\end{abstract}


\section{Introduction}
\IEEEPARstart{T}{here}  has been increasing interest in recent years towards applying ideas from coding theory to improve the performance of various computation \cite{HPC_gradient_2014, distrib_comp}, communication \cite{com, 6712144}, or algorithmic based fault tolerant solutions \cite{8723176}. A common ground for the majority of these research works is that they employ some form of real number parity based error correction code. Employing fixed-point or real numbers symbols instead of bits is naturally suited for execution on general purpose and digital signal processors (DSPs), that are well optimized for working with such data types, opposed to processing individual bits. The downside of real/fixed-point number codes is numerical stability due to quantization induced errors. However, as we will show, these shortcomings can be mitigated by quantization analysis and the choice of error correction code, as well as the encoding and decoding algorithms. Another possible downside is the additional bits required for representing the redundant symbols with respect to the data symbols. In a sense, this is the price to pay for their superior decoding performance with respect to Galois field counterparts which use modulo operations.\par

The idea of using  real-number codes that can be defined with generator matrices similar to those of the binary parity-check codes is not new \cite{turbo_analog, majority}. Neither is the usage of maximum likelihood (ML) decoding for real number codes \cite{turbo_analog, ML_conv}. However, to the best of our knowledge, this is the first attempt to propose an iterative gradient based solution for exact decoding of an implementation friendly subclass of real number codes inspired from binary quasi-cyclic LDPC codes. The proposed method is selected to limit the computational overhead of the encoder/decoder with respect to the binary counterparts by relying solely operations such as addition, comparison, counting, and maximum finder. \par
This paper is organized as follows: Section II discusses notations, the choice of parity check matrix, and encoding method, Section III presents the iterative gradient decoding, Section IV compares the decoding performance of the real domain with respect to the binary-domain, and provides a brief discussion of the implementation complexity and symbol quantization aspects for fixed point representation, while Section V provides some concluding remarks.


\section{A class of real-number QC-LDPC codes} \label{sec:linear_codes}
\subsection{QC-LDPC codes overview and notations} \label{sec:notations}
\gls{ldpc} codes are linear codes that can be described using a bipartite graph $\tilde{G}$, or alternatively using a sparse parity check matrix $H$ representation \cite{richardson2001capacity,richardson2001design,richardson}. This work uses \gls{qc-ldpc} codes that are a class of structured \gls{ldpc} codes that are obtained by \textit{expanding} a base matrix $B$ by an expansion factor. Specifically, each $B$ matrix entry is replaced by either a circulant permuted matrix, or by a zero matrix (denoted by $-1$, or $-$ in $B$) \cite{fossorier}. The non-negative entries from the base matrix $B$ are replaced by the unitary matrix shifted by the corresponding $B$ matrix entry value. The negative entries are replaced by the zero matrix. The parity check matrix $H$ will have size $(N_H - M_H) \times N_H$, with $N_H = z \times N_B$, $M_H = z \times M_B$, with $N_H$ representing the codeword size, $M_H$ the source bit/symbols vector($s$) length, $N_H - M_H$ the parity bits/symbols vector length, $z$ the expansion factor,  while $(N_B - M_B) \times M_B$ the size of $B$ matrix.   \par
\subsection{A subclass of QC-LDPC and its encoding} 
Without loss of generality, we selected LDPC codes that have $H$ in approximate lower triangular form. These codes are systematic, while the encoding process has low complexity. Furthermore, a wide range of LDPC codes for different communication standards, such as WiFi, WIMAX or 5G-NR, have this type of structure. Approximate lower triangular form can be obtained from all parity check matrices, via row and column permutations \cite{encoder_ru}; in this case, the source and parity bits/symbols are interleaved.
\IEEEpubidadjcol

\begin{figure}[!htb]
	\centering
	\begin{equation*}
	     \scalemath{0.98}{
		\begin{bmatrix}
			\begin{smallmatrix}
				25 & 26 & 14 & - & 20 & - & 2 & - & 4 & - & - & 8 & - & 16 & - & 18 & 1 & \textbf{0} & - & - & - & - & - & - \\ 
				10 & 9 & 15 & 11 & - & 0 & - & 1 & - & 1 & 18 & - & 8 & - & 10 & - & - & 0 & \textbf{0} & - & - & - & - & - \\ 
				16 & 2 & 20 & 26 & 21 & - & 6 & - & 1 & 26 & - & 7 & - & - & - & - & - & - & 0 & \textbf{0} & - & - & - & - \\ 
				10 & 13 & 5 & 0 & - & 3 & - & 7 & - & - & 26 & - & - & 13 & - & 16 & - & - & - & 0 & \textbf{0} & - & - & - \\ 
				23 & 14 & 24 & - & 12 & - & 19 & - & 17 & - & - & - & 20 & - & 21 & - & 0 & - & - & - & 0 & \textbf{0} & - & - \\ 
				6 & 22 & 9 & 20 & - & 25 & - & 17 & - & 8 & - & 14 & - & 18 & - & - & - & - & - & - & - & 0 & \textbf{0} & - \\ 
				14 & 23 & 21 & 11 & 20 & - & 24 & - & 18 & - & 19 & - & - & - & - & 22 & - & - & - & - & - & - & 0 & \textbf{0} \\ 
				17 & 11 & 11 & 20 & - & 21 & - & 26 & - & 3 & - & - & 18 & - & 26 & - & 1 & - & - & - & - & - & - & \textbf{0} \\ 
			\end{smallmatrix}
		\end{bmatrix}
	}
	\end{equation*}
	\caption{802.11 code rate 2/3 base matrix}
	\label{fig_code}
\end{figure}

For approximate lower triangular form matrices, we partition the base parity matrix into two sub-matrices, as shown in Fig. 1. Let $H = [H_s\; H_p]$ be the partitioned base parity check matrix, where $H_s$ is a matrix of size $(N_H-M_H) \times M_H$, corresponding to the source bits/symbols vector,  and $H_p$ a matrix of size $(N_H-M_H) \times (N_H-M_H)$, corresponding to the parity bits/symbols vector. For real number LDPC codes, we have modified the last diagonal in $H_p$ by replacing the 1 value from each expanded circulant identity matrix on the diagonal with -1.    
\par 
The encoding process is performed in a similar way to the binary version \cite{encoder1}, using a two step approach. The first set of $z$ parity symbols are obtained using \eqref{eq:enc_parity_1}.

\begin{subequations}\label{eq:enc_parity_1}
	\begin{align}
		\lambda_{i} &= H_p(i,:) \times s^{T}, i=1:{N_H-M_H} \\
		p(0,j) &= \sum_{k=1}^{N_H-M_H}\lambda_{k}(j), j=1:z
	\end{align}
\end{subequations}

The second step consists of obtaining the rest of $N_H - M_H - z$ parity symbols. Each set of $z$ parity symbols is derived using \eqref{eq:enc_parity_2}.

\begin{subequations}\label{eq:enc_parity_2}
	\begin{align}
		p(k,j) &= p(k-1,j) + \lambda_{k-1}(j), j=1:z
	\end{align}
\end{subequations}

\par
For the LDPC codes that have the parity check matrix having similar structure to the one depicted in Fig.~\ref{fig_code}, the encoding can be performed based on the parity check matrix and using only addition/subtraction operations.

\section{Decoding problem formulation}
Among possible decoding methods, the hard-decision algorithms such as Bit-Flip  \cite{bf}, or the gradient descent bit flipping (GDBF) algorithms proposed by Wadayama in \cite{gdbf}, are known low complexity algorithms, since symbols are represented by one bit. The binary decoders use the maximum likelihood (ML) decoding in the search for the codeword $x=(x_1, . . . , x_N)^T$ having the maximum correlation with the vector $\mathnormal{y}$ affected by noise. This letter investigates the changes and performance of a gradient descent decoding algorithm inspired from its binary counterpart the GDBF algorithm proposed by Wadayama. The operations during each iterations can be summarized as follows: \em{(1)} compute the active set of symbols for update as the set having the local energy $\ge$ a given threshold value, typically equal to the maximum local energy; \em{(2)} flip bits from active set.

The GDBF decoding method \cite{gdbf} defines an objective function, $E(x)$, also referred to as an energy function, and aims to maximize it. From it, the local energy function needed to compute the active set is derived. Furthermore, using an intuitive gradient descent method, the update operation is defined as bit-flipping of the active set symbols. Similar, for real number codes different formulations for the objective function have been proposed (see \cite{turbo_analog}). The formulation used in this letter is similar, with a key difference that stems from our goal, which is exact decoding. Thus, the metric we use is the number of symbols that are different at the end of decoding with respect to the reference instead of the mean square error.  Consequence of this, the energy function computation uses binarized variables in bipolar notation, similar to the binary counterparts. Let $s$ denote the syndrome vector computed as $s=Hx$, where $x$ refers to the current value of the symbol decode vector. A possible expression of the constraint function is: 
\begin{subequations}\label{eq:c1}
	\begin{align}
	E^{bin}_k &= \left(x_k-y_k\right)^{bin}+\sum_{\substack{m \in H(:,k) \\ H(m,k) \neq 0}}  s^{bin}_m \\
	D_k &= \sum_{\substack{m \in H(:,k) \\ H(m,k) \neq 0}}  sign(s_m) \\
	r &= (E^{bin}_1+\beta_1\times\left|D_1\right|, \dots, E^{bin}_N+\beta_N\times\left|D_N\right|) \\
	F^{bin}(x) &= \frac{1}{2}\left(\left\|r\right\|_{l_\infty}\right)^2 
	\end{align}
\end{subequations}
with $l_\infty$ being the infinity norm, $s^{bin}_m$ is the binarized syndrome value equal to $+1$ if the parity check is not satisfied, and $-1$ otherwise, while $sign(x)$ is the sign function. Note that, $E^{bin}_k$ has the same expression as for the binary case, and is a consequence of the exact decoding requirement. The additional term $D_k$ expresses whether the signs of all syndromes to which symbol $k$ contributes agree. If all parity checks are satisfied, $D_k$ is zero. It favors high column degree symbols. Furthermore, it is an indicator of the level of confidence in determining the sign of the error, and the direction for the error correction. The $\beta_k$ parameter is code dependent, having values in $\left[0,1\right]$. We aim to minimize $F^{bin}(x)$. The partial derivative is: \par
\begin{equation}\label{eq:lenergy}
	F^{local}_{k}=\frac{\partial F^{bin}}{\partial s_k} = \left(E^{bin}_k+\beta_k \times \left|D_k\right|)\right) \times \delta_{kj}
\end{equation} 
where $\delta_{kj}$ is the Kronecker delta equal to one for all cases where $ \left(E^{bin}_k+\beta_k \times \left|D_k\right|)\right)$ is equal to the maximum, and zero otherwise. (\ref{eq:lenergy}) suggests the active set of symbols indices needing correction as $F=\{n'|n' = arg \underset{k \in [1,N]}{max} F^{local}_{k}\}$. Since, ML decoding tries to determine the closest codeword, that means the smallest number of changed positions of $x$ with respect to $y$, and also the smallest possible amplitude for each correction of an arbitrary component $k$. In order to determine the direction of the correction for the approximate gradient descent method, we define the function $MajV(k)$ -- the majority voter function of the non-zero syndrome vector sign components, as:
\begin{equation}\label{eq:majSgn}
	MajV(k) = \twopartdef { sign(D_k) } {D_k \neq 0} {sign(y_k)} {D_k = 0},  k \in [1,N]
\end{equation}
For the magnitude, we select the minimum non-zero value, in accordance with the ML requirement. Thus, for the active set, the approximate correction for an arbitrary symbol $k$, denoted by $\delta_k$, is computed as:
\begin{equation}
	\begin{split}
		j = arg\;  \underset{\substack{m \in H(:,k) \\ H(m,k) \neq 0 \\ s_m \neq 0}}{min} |s_m| \\
		\delta_k= t_k \times MajV(k)\times|s_j|, t_k \in \mathbb{R}^+
	\end{split}
	\label{eq:update_value}
\end{equation}

Therefore, equation (\ref{eq:update_value})  defines the update rule for the real number codes, and $t_k$ is the update step factor.

\begin{algorithm}[] 
	\caption{Gradient Descent Real Number Symbol Update} \label{algo:real_gdbf}
	\begin{algorithmic}[1]
		\State{\textbf{Initialization:} } 
		\State{\text{    {\bf set} } $x_k=y_k$ for k=[1,N]  }

		\State{\text{{\bf compute}  $syndrome$ $s\leftarrow Hx$}}
		
		\While{$(i \le \textnormal{I}_{max})$ and $\left\|s\right\|_{l_1}=0$}
		\Comment{Iterative loop}
		\State{\textbf{Find symbol update position set $F$}}
		\State{compute active symbol set $F$}
		\ForAll{$x_k, k \in F$}
		\State{\textbf{compute $\delta_{k}$ acc. to (5)}}
		\State{$x_{k} \leftarrow x_{k} - \delta_{k}$}
		
		\EndFor
		\State{\text{{\bf compute}  $syndrome$ $s\leftarrow Hx$}}
		
		\State {{\bf set} $i \leftarrow i+1$}
		\EndWhile
		\State{\textbf{Offloading:} output $\gets$ $x$} 
	\end{algorithmic}
\end{algorithm}

The gradient descent algorithm for real number decoding is presented in Algo.~ \ref{algo:real_gdbf}. The $\left\|\dot{}\right\|_{l_1}$ is the $l_1$ norm, and $\textnormal{I}_{max}$ is the maximum allowed number of decoding iterations.

\section{Simulation Results and Analysis}
It has been demonstrated that any linear code defined over a finite field has a corresponding linear real-number code with similar error detecting and correcting capabilities \cite{RoundoffErrorsCodes_90}. The errors for real number codes can be regarded as Gaussian additive error values, statistically independent, with a given variance \cite{8723176}. Hence, the additive error model has two components: large amplitude errors, similar to impulse noise, and many small amplitude errors consequence of noise, or rounding errors that can be assimilated to white noise. The second category has been studied in \cite{RoundoffErrorsCodes_90, 8039530}, in the context of practical implementations that reply on finite precision fixed point numbers, and are more susceptible to round-off and overflow errors during the encoding, processing and decoding of the codewords. The present work addresses the first category. In this section, we analyze the decoding complexity and represent simulation results for real number codes derived from the binary IEEE 802.11n standard codes as described in Section II. \par 
For the reliability evaluation Monte Carlo simulations of up to 100 million input frames per simulation point have been performed. The channel model is the binary symmetric channel with crossover probability $\alpha$. The binary symbols are flipped, while the non-binary ({\em e.g.} real or fixed-point) symbols are added amplitude errors with $\alpha$ probability. The metric bit error rate (BER) for the binary codes, and symbol error rate for the non-binary (SER) account for the number of erroneous decoder output bits, and non-binary symbols respectively. Similarly, we use the frame error rate (FER), and the output error rate, to report the ratio of failed output frames for the binary, and non-binary case, out of the total simulated frames for a channel parameter $\alpha$. The parameters $\beta_i$ and $t_i$ are set to 1. As depicted in Figs.2-a,3, this choice is favorable for the rate $1/2$ code, yielding two order of magnitude decoding performance improvement of the real case with respect to the binary counterpart, one order of magnitude for the rate $2/3$ code, and roughly the same performance for the $3/4$ rate code.  For the average iteration, the charts show a similar trend as for the rate $1/2$ code (Fig.2-b).\par 

Without loss of generality, for the complexity analysis we consider the case of $p$ bit  fixed point symbols, and regular codes with column degree ({\em i.e.} number of ones in a parity check matrix column) denoted by $d_v$, and row degree ({\em i.e.} number of ones in a parity check matrix row) denoted by $d_c$. We compare in Table~\ref{tab:complexity} the number and type of operations required for computing the binary and the real number gradient decoding. It can be seen that the additional overhead is limited to the replacement of the XOR operations from Galois Field 2 with additions. The additional precision bits for the redundant symbols for fixed point arithmetic considering an adder tree implementation is $\lceil d_c \rceil$. Therefore, the encoder and the parity check computations need to work with worst adders on $p+\lceil d_c \rceil$ precision bits. Also, all redundant symbols require $p+\lceil d_c \rceil$+1 bit precision in order to guarantee numerical stability. If this requirement is met the real number simulation curves and the fixed point simulation curves have similar the same performance.
 
 \begin{figure*}
 	\centering
 	\subfloat[]{\includegraphics[width=3.5in]{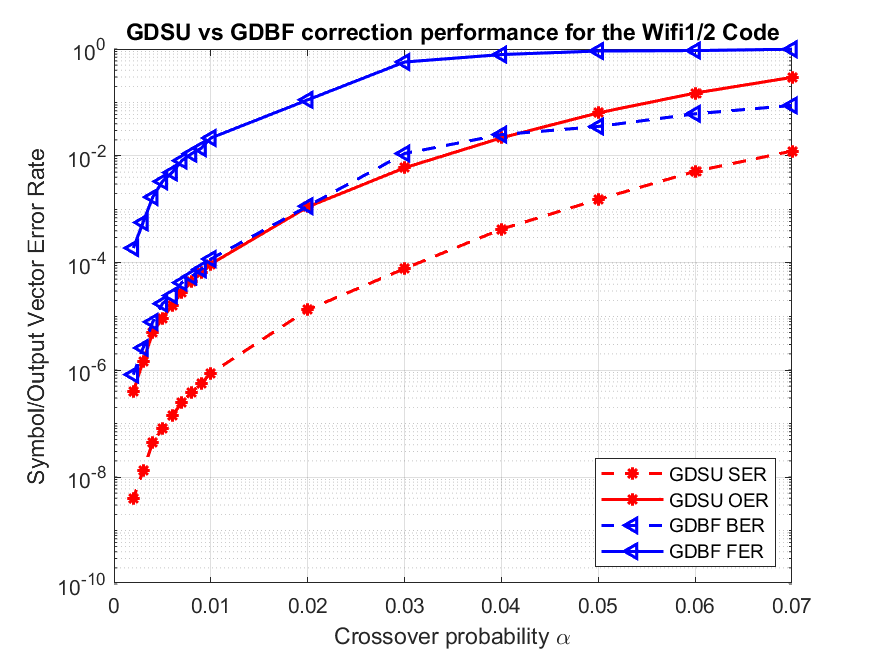}%
 		\label{fig_FER_1_2}}
 	\hfil
 	\subfloat[]{\includegraphics[width=3.5in]{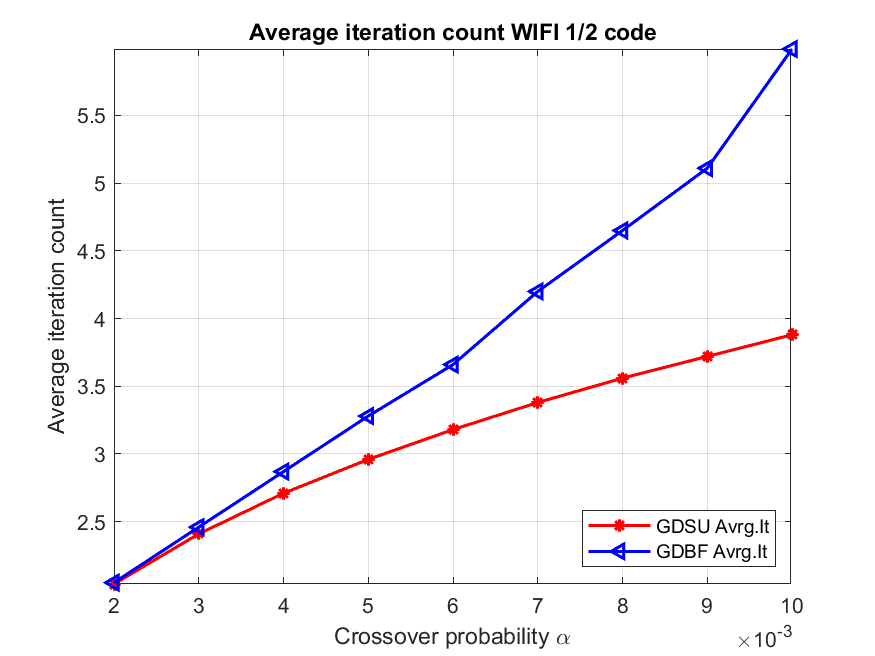}%
 		\label{fig_AvrgIt_1_2}}
 	\caption{Wifi $1/2$ code rate for the binary (GDBF) and non-binary (GDSU) cases (a) Decoding performance (b) Average number of decoding iterations.}
 	\label{fig_sim}
 \end{figure*}

\begin{figure}[htb]
	\centering
	\includegraphics[width=3.2in]{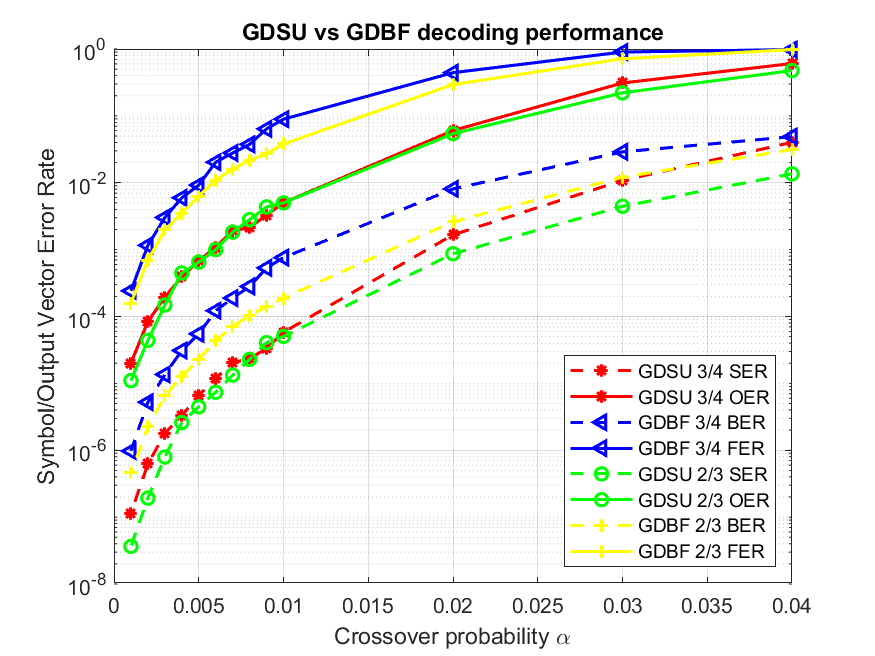}
	\caption{Decoding performance for rate 2/3 and 3/4 codes}
	\label{fig_1}
\end{figure}

\begin{table}[!tb] \label{tab:complexity}
	\caption{Complexity of implementing binary and real number decoding for fixed point numbers using $p$ bits precision per symbol }
	\centering
	\begin{tabular}{|c||c|c|}
		\hline
		Decoding & Binary & Real \\
		\hline
		$s$ computation  &  $M \times d_c$ XOR ops& $M \times d_c$ ADD ops \\
		\hline
		Local energies & $N \times d_v$ ADD ops & $2\times N \times d_v$ ADD ops\\
		\hline
		Max. finder & $N \times (N-1)$ comp.  & $N \times (N-1)$ comp.\\
		\hline
		Symbol update & 1 XOR op per symbol & 1 ADD op per symbol\\
		\hline

	\end{tabular}
\end{table}

\section{Conclusion}

This paper proposed a novel iterative gradient descent decoding method for real number parity codes, and evaluated the decoding performance and average iteration count for a sub-class of QC-LDPC codes inspired by the binary LDPC  communication standard codes. The encoding and the decoding of QC-LDPC codes investigated in this paper relies solely on additions based operations. This is both cost efficient, and also simplifies the numerical stability analysis. For the rate $1/2$ 802.11 code, the GDSU outperforms the binary GDBF, while for higher rates it exhibits slightly higher, or similar performance as the GDBF. One possible reason for this behavior is the fact that the $\beta_i$ and $t_k$ parameters have not been properly investigated. They have been set to a default value of one. This will be the subject of future work.



\bibliographystyle{IEEEtran}
\bibliography{biblio_database}



\vfill

\end{document}